\documentclass{aastex61}

\usepackage{natbib,amsmath}
\shorttitle{XMM-Newton Spectroscopy of Two Red Quasars} 
 \shortauthors{Glikman et al.}
 
\begin{document}

\title{Peering through the dust II: {\em XMM$-$Newton} observations of two additional FIRST-2MASS red quasars}

\author{Eilat Glikman}
\affiliation{Department of Physics, Middlebury College, Middlebury, VT 05753, USA}

\author{Stephanie LaMassa}
\affiliation{Space Telescope Science Institute, 3700 San Martin Drive, Baltimore MD, 21218, USA}
\affiliation{NASA Goddard Space Flight Center, Greenbelt, MD 20771, USA}

\author{Enrico Piconcelli}
\affiliation{Osservatorio Astronomico di Roma (INAF), via Frascati 33, 00040 Monte Porzio Catone (Roma), Italy}

\author{Meg Urry}
\affiliation{Yale Center for Astronomy \& Astrophysics, Physics Department, P.O. Box 208120, New Haven, CT 06520, USA}
\affiliation{Department of Physics, Yale University, P.O. Box 208121, New Haven, CT 06520, USA}

\author{Mark Lacy}
\affiliation{National Radio Astronomy Observatory, 520 Edgemont Road, Charlottesville, VA}

\email{eglikman@middlebury.edu}

\begin{abstract}
We obtained {\em XMM-Newton} observations of two highly luminous dust-reddened quasars, F2M1113+1244 and F2M1656+3821, that appear to be in the early, transitional phase predicted by merger-driven models of quasar/galaxy co-evolution. These sources have been well-studied at optical through mid-infrared wavelengths and are growing relatively rapidly, with Eddington ratios $>30\%$. Their black hole masses are relatively small compared to their host galaxies placing them below the $M_{\rm BH} - L_{\rm bulge}$ relation. We find that for both sources, an absorbed power-law model with $1-3\%$ of the intrinsic continuum scattered or leaked back into the line-of-sight best fits their X-ray spectra.  
We measure the absorbing column density ($N_H$) and constrain the dust-to-gas ratios in these systems, finding that they lie well below the Galactic value.  This, combined with the presence of broad emission lines in their optical and near-infrared spectra, suggests that the dust absorption occurs far from the nucleus, in the host galaxy, while the X-rays are mostly absorbed in the nuclear, dust-free region within the sublimation radius.  
We also compare the quasars' absorption-corrected, rest-frame X-ray luminosities ($2-10$ keV) to their rest-frame infrared luminosities (6$\mu$m) and find that red quasars, similar to other populations of luminous obscured quasars, are either underluminous in X-rays or overluminous in the infrared. 
\end{abstract}


\section{Introduction}

A complete picture of galaxy evolution must include a prescription for the growth of supermassive black holes (SMBHs) at their centers, as SMBHs are ubiquitous in the nuclei of galaxies \citep{Faber97} and tight correlations are seen between their masses and host galaxy properties, e.g., the  $M_{\rm BH} - \sigma$ relation \citep{Gebhardt00,Ferrarese00} and the $M_{\rm BH} - L_{\rm bulge}, M_{\rm bulge}$ relations \citep{Marconi03,Haring04}.
These observations can be explained by models of major galaxy mergers that trigger both SMBH accretion and circumnuclear star-formation \citep[e.g.,][]{Hopkins06a,Hopkins06c}. According to that scenario, the resultant obscuring dust is eventually cleared by powerful quasar winds, revealing luminous, unreddened emission from the quasar (the scenario originally described by Sanders et al. 1988). In this scenario dust-reddened (or ``red'') quasars appear to represent a key phase in SMBH/galaxy co-evolution: the transition from a dust-enshrouded core, likely surrounded by a nuclear starburst, to a typical unobscured blue quasar stage \citep{Glikman12}. 

We have identified a sample of luminous red quasars in a radio-selected survey that combined FIRST \citep{White97} and 2MASS \citep{Skrutskie06} with follow-up spectroscopy at optical and near-infrared wavelengths \citep[F2M;][]{Glikman04,Glikman07b,Glikman12}. This complete sample of 120 red quasars spans a broad range of redshifts $(0.1 < z < 3)$ and reddenings $(0.1 < E(B - V ) < 1.5)$. With this large sample, \citet{Glikman12} conducted a systematic study of red quasar properties as a function of redshift, luminosity and reddening, concluding that quasars remain at least partly shrouded during their most luminous phase.  This reddened phase -- presumably occurring after the quasar has become powerful enough to blow away some of the circumnuclear obscuration -- therefore represents a very important component of SMBH growth. 

{\em Hubble Space Telescope} ({\em HST}) imaging of F2M red quasar hosts reveals an usually high fraction $(>80\%)$ of mergers and/or interacting systems at $z \sim 0.7$ \citep{Urrutia08} and $z\sim2$ \citep{Glikman15}.
We also see a high fraction $(> 60\%)$ of low ionization broad absorption line quasars \citep[LoBALs and FeLoBALs, which make up $\lesssim 2\%$ of normal quasars,][]{Trump06} indicative of outflows associated with feedback \citep{Urrutia09,Farrah12,Glikman12}.

{\em Spitzer} observations of the same thirteen red quasars studied in \citet{Urrutia08} enabled full modeling of their spectral energy distributions (SEDs) and a determination of their bolometric luminosities. \citet{Urrutia12} combined their bolometric luminosities with the broad lines in their spectra to measure black hole masses; they find that most of these objects are accreting at high Eddington rates $(L/L_{\rm Edd} > 0.3)$. 
Furthermore, the systems with the highest $L/L_{\rm Edd}$ lie below the $M_{\rm BH} - L_{\rm bulge}$ relation, implying that these black holes have yet to grow to to their equilibrium size following a major merger.

While multi-wavelength observations, from radio through optical, paint a compelling picture of dust-reddened quasars as the clearing phase in quasar/galaxy co-evolution driven by feedback processes, they have not been well studied in X-rays, which can provide independent checks on these conclusions.  
Since X-rays are able to pierce through obscuring dust, X-ray observations can give a measure of the true underlying accretion luminosity, for all but the most obscured AGN \citep[$N_H > 10^{24}$ cm$^{-2}$;][]{Comastri04}.  
We can compare optical and X-ray measured reddening to determine the average gas-to-dust ratio \citep[which may be anomalous in AGN;][]{Maiolino01}.  Additionally, the slope of the X-ray power law (i.e., the photon index, $\Gamma$, where $L(E) \propto E^{-\Gamma}$) may provide insight into the SMBH accretion rate \citep{Shemmer06,Shemmer08,Brightman16}, though recent analysis of nearby, hard X-ray-selected AGN disputes this \citep{Trakhtenbrot17}.

A handful of F2M red quasars have been observed with the {\em Chandra} and  {\em XMM-Newton} X-ray observatories, but not for sufficient exposure times to provide detailed spectra, which are necessary to derive realistic values for $N_H$ and $\Gamma$.   
Twelve F2M quasars were observed with {\em Chandra} ACIS-S in 2004. 
Most of the objects were only weakly detected in $5-10$ ksec providing hardness ratio (HR) measurements indicative of moderate absorption \citep{Urrutia05}.  
\citet{Wilkes02,Wilkes05} studied the X-ray properties of 2MASS-selected red AGN \citep{Cutri01} and found similar results, emphasizing the objects' faintness and the heterogeneous nature of their absorption properties. 

More recently, \citet{LaMassa16} presented an analysis of two F2M red quasars observed with {\em NuSTAR}, F2M0830+3759 and F2M1227+3214, previously detected in X-rays with {\em Chandra} and {\em XMM}. F2M0830+3759, at $z=0.414$, was part of the \citet{Urrutia05} study and was subsequently observed for 50 ksec with {\em XMM} \citep{Piconcelli10}.  It was also among the thirteen red quasars studied with {\em HST} \citep{Urrutia08} and {\em Spitzer} \citep{Urrutia12} and was found to have a merging host galaxy, high accretion rate, and small black hole mass (relative to its host galaxy's bulge luminosity).  In their analysis of the {\em XMM} observation, \citet{Piconcelli10} suggest that the reddening and absorber likely occur in the same medium, with the X-ray soft-excess likely due to a combination of emission from scattered continuum photons and distant photoionized gas, implying the presence of material off the line-of-sight.

F2M1227+3214, $z=0.137$, was observed with {\em Chandra} for 3.7 ksec and yielded a spectrum with $\sim 800$ counts.  This source also has a high accretion rate, but with no high-resolution imaging available, its host morphology is unknown.

F2M0830+3759 was observed with {\em NuSTAR} for 22 ksec and F2M1227+3214 was observed for 23 ksec.  Both observations yielded $\gtrsim 1000$ counts in the $3-79$ keV energy range.  \citet{LaMassa16} performed detailed spectral analysis of these sources, utilizing both phenomenological absorbed-power-law models as well as self-consistent physically-motivated models such as BNTorus \citep{Brightman11} and MYTorus \citep{Murphy09}.  While F2M1227+3214 was well-described by a lightly-absorbed power law with $N_H = 3.4\times10^{21}$ cm$^{-2}$, F2M0830+3759 warranted more complex modeling.  The best interpretation of this source's X-ray spectrum is a patchy obscuring medium surrounding a power-law source, resulting in an absorbed power law for photons traveling along the line of sight whose absorption is due to a column density of $N_H = 2.1\times10^{22}$ cm$^{-2}$ plus scattered emission from ``global'' obscuring material out of the line of sight whose column density is much higher, $N_H = 3.7\times10^{23}$ cm$^{-2}$.  
Consistent with the interpretation of \citet{Piconcelli10}, the {\em NuSTAR} modeling is suggestive of a galaxy-scale absorber as expected from its disturbed morphology.  

In this paper we present deep {\em XMM-Newton} observations of two additional red quasars selected from the subsample of objects that have high-resolution imaging from {\em HST} as well as infrared photometry and spectroscopy out to 160\micron\ from {\em Spitzer}.  Similar to F2M0830+3759 described above, F2M1113+1244 and F2M1656+3821 have accretion rates $> 30\%$ of the Eddington limit and lie below the $M_{BH} - L_{\rm bulge}$ relation \citep{Urrutia12}.   Table \ref{tab:objects} lists their general properties, along with the two other red quasars from \citet{LaMassa16}.  




\begin{deluxetable}{ccccccc}



\tablewidth{0pt}
 
\tablecaption{Red Quasar Properties\label{tab:objects}}

\tablenum{1}

\tablehead{\colhead{Name} & \colhead{R.A.} & \colhead{Dec} & \colhead{Redshift} & \colhead{$E(B-V)$} & \colhead{$\log(M_{\rm BH}/M_\odot)$\tablenotemark{a}} & \colhead{$\log(L/L_{\rm Edd})$\tablenotemark{a}} \\ 
\colhead{} & \colhead{(J2000)} & \colhead{(J2000)} & \colhead{} & \colhead{(mag)} & \colhead{} & \colhead{} } 

\startdata
F2M0830+3506 &  08:30:11.12 &  +37:59:51.8 &  0.414 &  0.73\tablenotemark{b} &   $8.6\pm0.1$ &  $-0.4\pm0.1$ \\
{\bf F2M1113+1244} & 11:13:54.67 &  +12:44:38.9 &  0.681 &  1.51\tablenotemark{b} &   $8.6\pm0.1$ &  $0.4\pm0.1$ \\
F2M1227+3214 &  12:27:49.15 &  +32:14:59.0 &  0.137 &  0.71\tablenotemark{c} &   $7.4\pm0.1$ &  $-0.1\pm0.1$ \\
{\bf F2M1656+3821} &  16:56:47.11 &  +38:21:36.7 &  0.732 &  0.61\tablenotemark{b} &   $8.5\pm0.1$ &  $-0.2\pm0.1$ \\
\enddata

\tablecomments{Objects listed in boldface type are the subjects of the present paper. }
\tablenotetext{a}{Black hole mass values and Eddington ratios are taken from \citet{Urrutia12} for all sources except F2M1227+3214, for which we computed the mass following the same procedure. }
\tablenotetext{b}{Reddening values as reported in \citet{Glikman12}} 
\tablenotetext{c}{Reddening values as reported in \citet{LaMassa16}}



\end{deluxetable}

In this paper we analyze the X-ray spectra of these two red quasars and, together with the previous {\em NuSTAR}-observed sample, examine their accretion and obscuration properties.
Throughout this paper, we adopt the $\Lambda$CDM concordance cosmology: $H_0=70$ km s$^{-1}$ Mpc$^{-1}$, $\Omega_M=0.30$, and $\Omega_\Lambda=0.70$ when computing cosmology-dependent quantities \citep{Bennett13}.

\section{Multi-wavelength Observations}

\begin{figure}
\plottwo{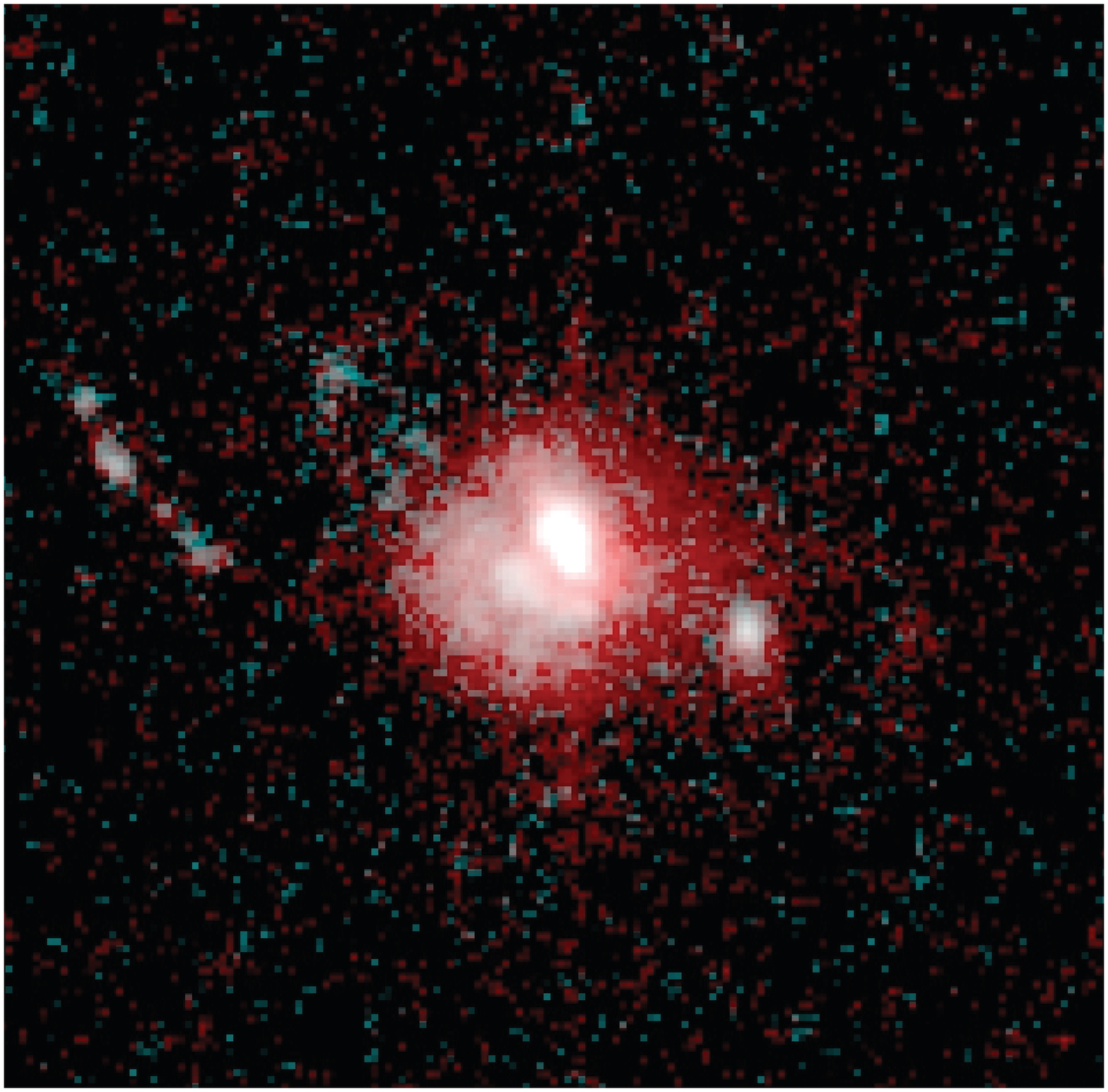}{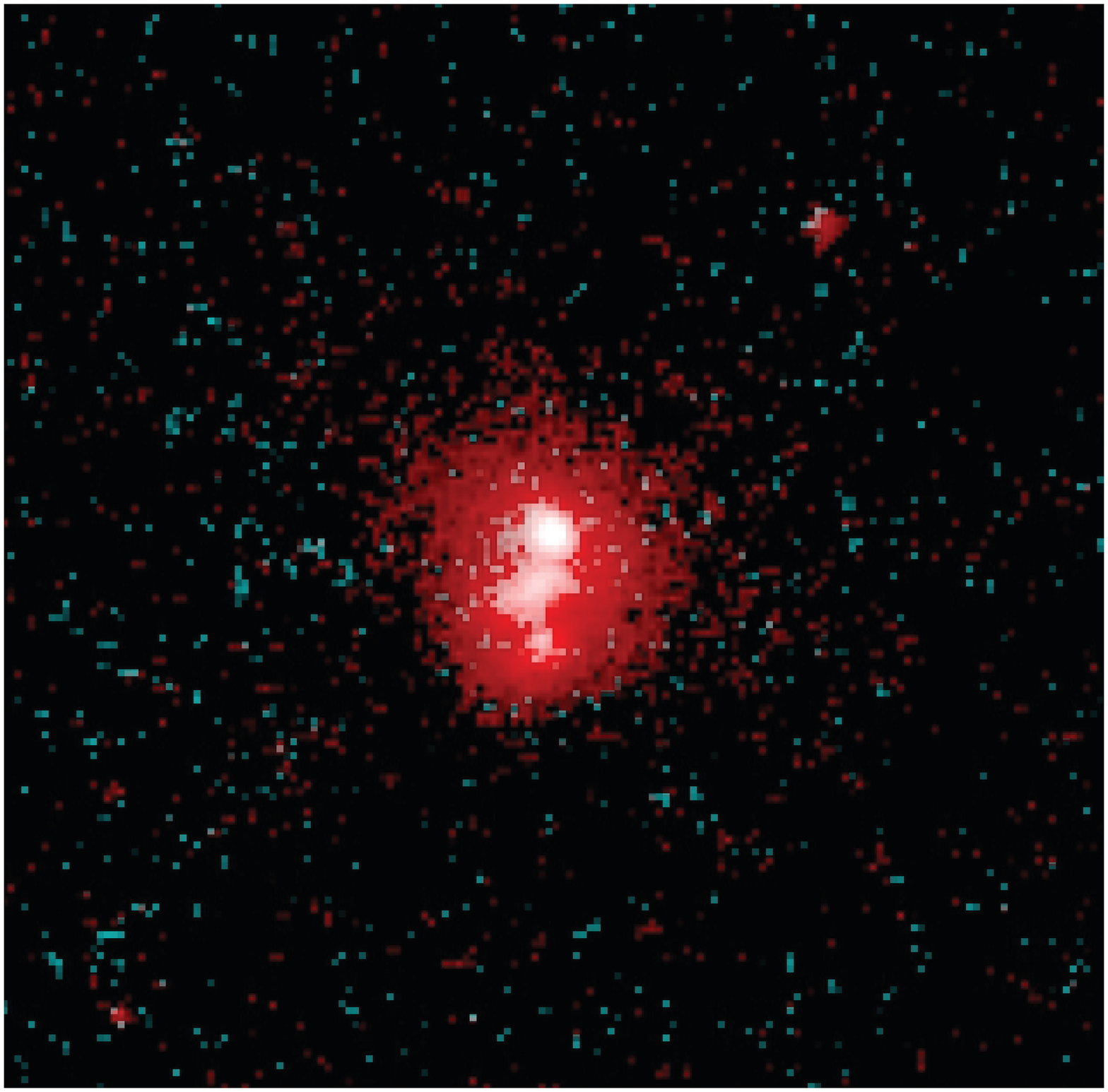}
\caption{Red-Green-Blue (RGB) color composites of the {\em HST} images of the two red quasars in this study. F2M1113+1224 is shown in the left panel, and F2M1656+3821 is shown on the right.  We used the F814W band for the red and the F475W band for the blue and green parts of the composite. Both images show multi-component, asymmetric sources, highly suggestive of host-galaxy mergers with the brightest source in each image representing the reddened quasar. Cutout sizes are $8\arcsec \times 8\arcsec$.}\label{fig:hst}
\end{figure}

\begin{figure}
\plotone{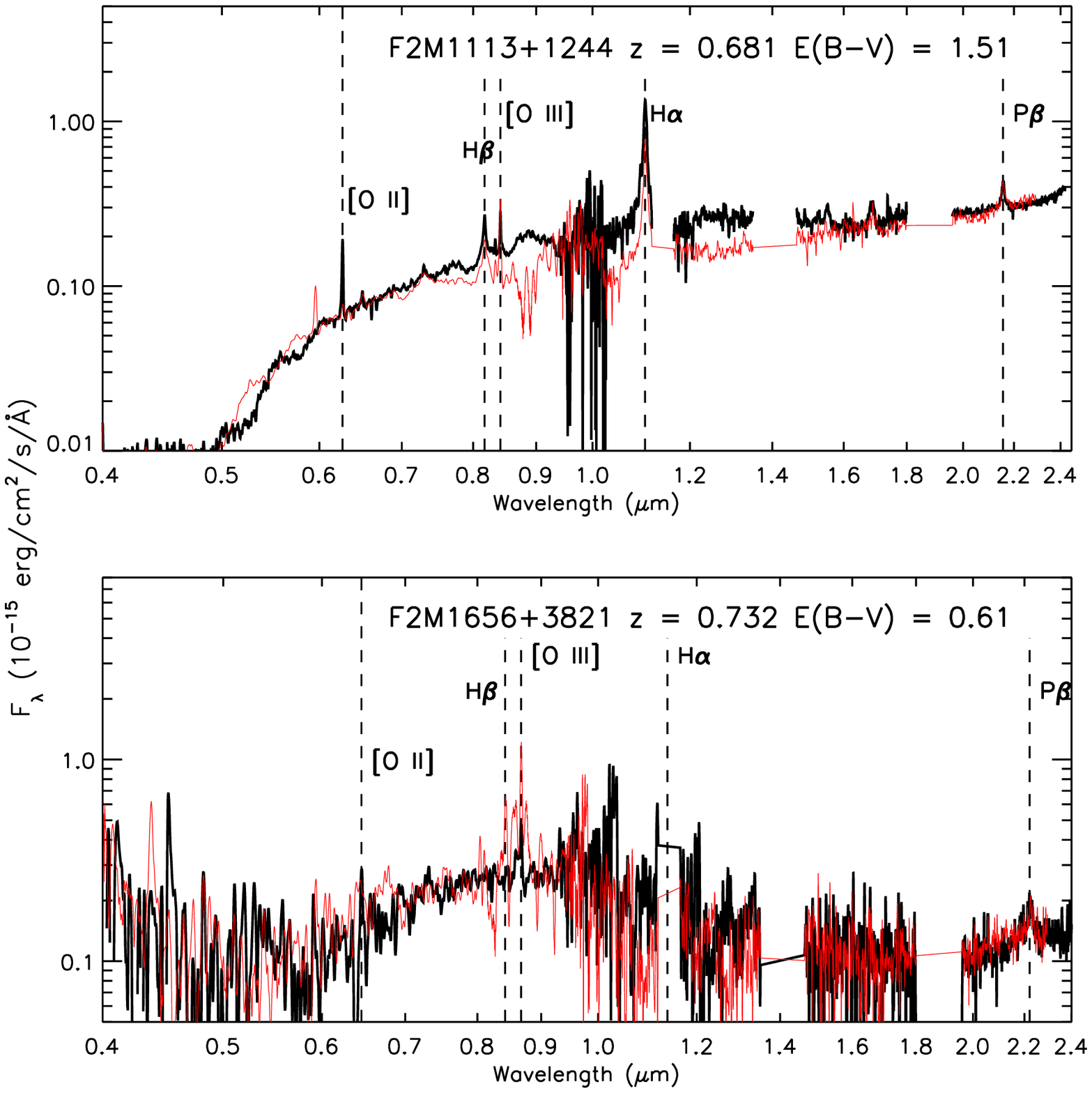}
\caption{Optical through near-infrared spectra (black lines) of F2M1113+1244 (top panel) and F2M1656+3821 (bottom panel).  The red line is the best-fit reddened quasar template, following the fitting method described in \citet{Glikman07} to find the color-excess, $E(B-V)$.}\label{fig:ebv}
\end{figure}

F2M1113+1244 and F2M1656+3821 are two well-studied red quasars with photometry spanning $\sim 3500$ \AA\ through $>70~\mu$m, spectroscopy at optical and near-infrared wavelengths, and high-resolution optical imaging with {\em HST}.  Figure \ref{fig:hst} shows their color-combined {\em HST} images, which clearly show their asymmetric and disturbed morphologies.  In Figure \ref{fig:ebv} we plot their optical through near-infrared spectra overlaying a fit to a reddened quasar template, from which we derive their extinction properties, $E(B-V)$, listed in Table \ref{tab:objects}.  

The red quasars in the F2M survey required a detection in the FIRST radio catalog and thus have a 20 cm flux density brighter than 1 mJy.  \citet{Glikman12} examined the effect of radio luminosity on reddening in the overall sample and found no relation between reddening and radio luminosity.  However, radio jets may affect the X-ray emission \citep{Miller11}.  We examine the radio-loudness, $R$, of these two quasars applying the the definition of \citet{Ivezic02}, $R= \log(F_{\rm radio}/F_{\rm optical})=0.4(g_{\rm cor} - t)$, where $g_{\rm cor}$ is the reddening-corrected SDSS $g$-band magnitude and $t = -2.5\log(F_{\rm int}/ 3631 {\rm Jy})$ is a magnitude-equivalent radio flux\footnote{This logarithmic definition of radio loudness differs from the definition of \citet{Stocke92}, which is a linear quantity.}. A radio-quiet quasar is defined as having $R\lesssim0.5$ and a radio-loud quasar has $R\ge2$.  Quasars with $R$ lying between these values are considered ``radio-intermediate''. We find that F2M1113+1244 has $R=0.68$ and F2M1656+3821 has $R=2.47$.  The {\em NuSTAR}-observed red quasars from \citet{LaMassa16} have $R=1.14$ (F2M0830+3506) and $R=0.35$ (F2M1227+3214).  Thus, of the quasars studied here, only F2M1656+3821, with a FIRST flux density of 3.9 mJy, is radio loud.

\section{X-Ray Observations and Reduction}

{\em XMM-Newton} observed F2M1113+1244 for 28 ksec on UT 2015 May 21 and F2M1656+3821 for 81 ksec on UT 2016 March 17-18. 
We retrieved the ODF files for our observations and reduced them using Science Analysis Software (SAS) version 15.0.0 and HEASOFT v.6.19, following the SAS reduction threads related to the European Photon Imaging Camera \citep[EPIC;][]{Struder01,Turner01}.  We filtered the events file for flaring, keeping time intervals whose count rates fell below 0.35 for the MOS detectors and 0.4 for the PN detector.  The effective MOS detector exposures become 28 ksec and 58 ksec for F2M1113+1244 and F2M1656+3821, respectively, and the effective PN detector exposures become 24 ksec and 29 ksec, respectively.  Table \ref{tab:obssum} summarizes the details of our observations. 




\begin{deluxetable}{cccccccccc}



\tablewidth{0pt}

\tablecaption{Summary of {\em XMM-Newton} Observations \label{tab:obssum} }

\tablenum{2}

\tablehead{\colhead{} & \colhead{} & \colhead{} & \colhead{} & \multicolumn{3}{c}{Net Exposure Time} & \multicolumn{3}{c}{Net Count-rate ($\times 10^{-3}$)}\\
\colhead{Name} & \colhead{ObsID} & \colhead{Date} & \colhead{$N_{H,{\rm Galactic} }$\tablenotemark{a}} & \colhead{PN} & \colhead{MOS1} & \colhead{MOS2} & \colhead{PN} & \colhead{MOS1} & \colhead{MOS2} \\ 
\colhead{} & \colhead{} & \colhead{} & \colhead{(cm$^{-2}$)} & \colhead{(ksec)} & \colhead{(ksec)} & \colhead{(ksec)} & \colhead{(cnts s$^{-1}$)} & \colhead{(cnts s$^{-1}$)} & \colhead{(cnts s$^{-1}$)} } 

\startdata
F2M1113+1244 &  0761950101 &  2015 May 21&    $1.82\times 10^{20}$ &  23.58 &    27.74 &  27.73 &  $18\pm 1$ &  $5.1\pm0.5$ &  $5.2\pm0.6$ \\
F2M1656+3821 &  0761950201 &  2016 March 17$-$18 &  $1.93\times 10^{20}$ &  29.05 &    58.89 &  58.32 &  $29\pm 1$ &  $9.3\pm0.6$ &  $9.2\pm0.5$ \\
\enddata

\tablenotetext{a}{Galactic absorption determined from the maps of \citet{Dickey90}.}



\end{deluxetable}

We extracted spectra from the MOS detectors using an aperture with a 35\arcsec\ radius centered on both targets and extracted a background region from an annulus with an inner radius of 45\arcsec\ and outer radius of 85\arcsec. For the observations of F2M1113+1244 with the PN detector, due to the source's proximity to the chip gap, we used an aperture with a 30\arcsec\ radius, and selected several source-free regions surrounding the source, whose area was similar to the annular region.   F2M1656+3821 was also near the chip gap, but the 35\arcsec-radius was possible.  
 After filtering and extraction, the final counts for F2M1113+1244 are $\sim 700$ ($\sim 415$ in PN and $\sim 140$ in each MOS detector) and the counts for F2M1656+3821 are $\sim 1900$ ($\sim 840$ in PN and $\sim 540$ in each MOS detector), respectively.  
 We  group the spectra into bins of 5 counts using the HEASOFT command {\tt grppha}, using the Cash statistic in our fitting \citep[C-stat;][]{Cash79}.

\subsection{Absorbed Power Law Models}

We use the X-ray fitting software {\em XSpec} v12.9.0 \citep{Arnaud96} to investigate the X-ray emitting nature of our quasars, performing a simultaneous fit to the three spectra from the different detectors, in order to utilize all the counts. 
We began by fitting a simple power-law fit, {\tt phabs*zpow}, to both sources allowing only absorption in the Milky Way ({\tt phabs}). Table \ref{tab:obssum} lists the Galactic hydrogen column density along the line-of-sight to each quasar,\footnote{From {\tt cxc.harvard.edu/toolkit/colden.jsp} using the maps of \citet{Dickey90}.} which we fix in all our fits.
Neither yields a satisfactory fit, with F2M1113+1244 having a best-fit photon index of $\Gamma=0.344$ (C-stat 254.25 for 161 DOF) and F2M1656+3821 having $\Gamma=0.279$ (C-stat 764.24 for 499 DOF).  Neither of these photon indices are close to the typical intrinsic AGN value of $\Gamma \simeq 1.8$.  
We next tried an absorbed-power-law model, {\tt constant*phabs*zphabs*zpowerlw}, with absorption occurring both at the source ({\tt zphabs}) and in the Milky Way to accommodate the flattening of the spectrum.  

Although moderate absorption is found for both sources with this simple model ($N_H \sim 10^{22}$ cm$^{2}$) we see a distinct excess at soft energies below 2 keV.  This suggests that there may be scattered or leaked light at lower energies\footnote{Note that the {\em XSpec} modeling is unable to distinguish between whether the soft emission is coming from X-rays that are scattered through an opening in the ``torus'' and then off an optically-thin medium into our line-of-sight or whether the soft emission is leaking through a patchy medium to enter our line-of-sight directly. Both processes would have the same effect on the spectrum, and are thus described by the same combination of power-laws in {\em XSpec}.} in excess of the absorbed primary continuum.  Thus we try a double-absorbed power law with the same photon index for both components, which is effectively a partial covering model, 

\begin{equation}
\begin{split}
model = & ~C \times \exp[-N_{\rm H,Gal}\sigma(E)] \\
           & \times \Big(A_1 \times E_r^{-\Gamma} + \exp[-N_{\rm H,Z}\sigma(E_r)]\times A_2\times E_r^{-\Gamma}\Big), \label{eqn:dpl}
\end{split}
\end{equation}
where $E_r$ is the rest-frame energy emitted by the source, i.e., $E_r = E[1+z]$ with $E$ as the observed energy. 
In XSpec parlance the model is {\tt constant*phabs*(zpowerlw + zphabs*zpowerlw)}. 

\begin{figure*}
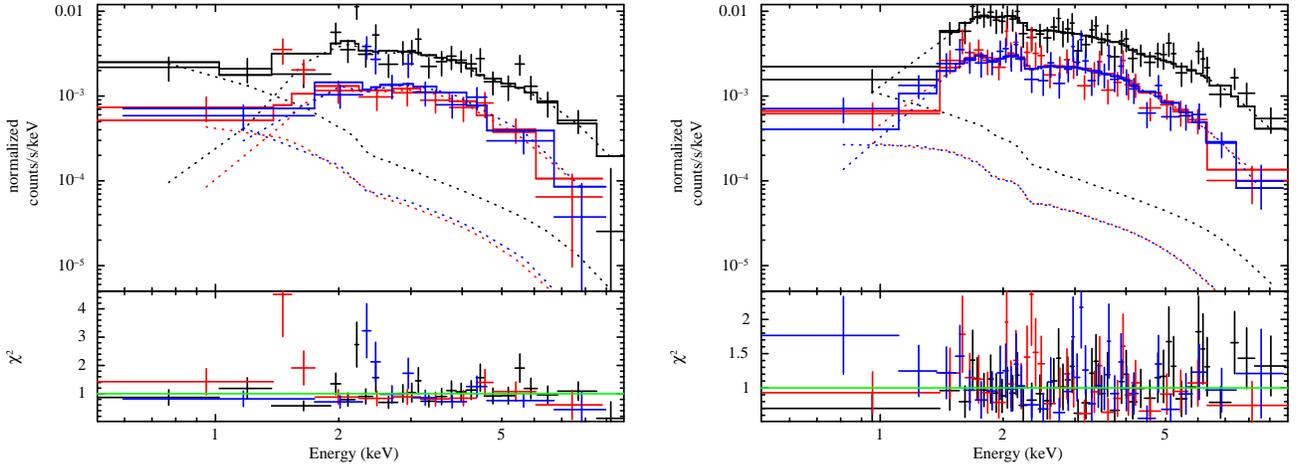

\includegraphics[angle=270,scale=0.35]{f3a.ps}
\includegraphics[angle=270,scale=0.35]{f3b.ps}
\caption{Best-fit partial covering absorbed power law (Equation \ref{eqn:dpl}) fits to the {\em XMM} quasar spectra in the energy range  $0.5 - 10$ keV. {\em Left --} F2M1113+1244. {\em Right --} F2M1656+3821.
Dotted lines represent the individual components. The solid line shows the combined model. Red and blue points and lines are for data from the MOS detectors. Black points and lines are for data from the PN detector.  Bottom panels show the residuals.  In both fits we find that the sources are heavily absorbed ($N_H \sim 10^{23}$ cm$^{-2}$) with a small amount of leaked flux ($\sim 1-3\%$).}\label{fig:xspecdpl}
\end{figure*}

Adding an unabsorbed power-law component improves the fit.  
The test statistic for the fit to F2M1113+1244 becomes C-stat =165.91 for 157 degrees of freedom (from 193.46 for 159 dof) and the best fit parameters are: $\Gamma=2.06$, $N_H = 1.3\times10^{23}$ cm$^{-2}$, and 2.8\% of the source radiation leaked into the unabsorbed component. 
The test statistic for F2M1656+3821 is C-stat = 499.33 for 495 degrees of freedom (from 508.68 for 497 dof), with $\Gamma=1.81$, $N_H = 9.2\times10^{22}$ cm$^{-2}$, and 1.5\% of leaked radiation.  
We note that these fits are robust against the choice of binning, and give the same results when we fit data binned by 10 and 20 counts.  
The fitted spectra are shown in Figure \ref{fig:xspecdpl}. 
Table \ref{tab:param} reports the fitted parameters for this model for both sources.  The fraction of leaked flux is found by taking the ratio of $A_1 / A_2$ and reported under the column headed $f_{\rm scatt}$.  

We check to see whether the soft X-ray excess is due to star formation rate by computing the luminosity of scattered component in the soft band ($0.5 - 2$ keV) and converting it to a star formation rate using the relation of \citet{Ranalli03}.  We find the scattered luminosity for F2M1113+1244 to be $L_{\rm soft} = 3.2\times 10^{44}$ erg s$^{-1}$ which translates into a star formation rate of 70,400 $M_\sun$ yr$^{-1}$. For F2M1656+3821 we find $L_{\rm soft} = 2.9\times 10^{44}$ erg s$^{-1}$ which translates into a star formation rate of 63,800 $M_\sun$ yr$^{-1}$. Both these values are very extreme and inconsistent with their infrared emission and SED fits reported in \citet{Urrutia12} which estimate star formation rates of a few tens $M_\sun$ yr$^{-1}$. We thus do not consider star-formation to contribute significantly to the soft X-ray emission.




\begin{deluxetable}{ccccccc}



 \tablewidth{0pt}
 
\tablecaption{Best  Fits to X-ray Spectra\label{tab:param}}

\tablenum{3}

\tablehead{\colhead{Name} & \colhead{Model} & \colhead{$\Gamma$} & \colhead{$N_{\rm H,Z}$} & \colhead{$f_{\rm scatt}$} & \colhead{$\theta_{\rm obs}$} & \colhead{C-stat (dof)} \\ 
\colhead{} & \colhead{Equation} &  \colhead{} & \colhead{($10^{23}$ cm$^{-2}$)} & \colhead{(\%)} & \colhead{(degrees)} & \colhead{} } 

\startdata
F2M1113+1244 & \ref{eqn:dpl}     & $2.07^{+0.36}_{-0.35}$ &  $1.34^{+0.34}_{-0.31}$ &   $2.6^{+1.7}_{-1.1}$ &      \ldots &      165.91 (157) \\
-            & \ref{eqn:mytorus} & $2.06^{+0.37}_{-0.35}$ &  $1.30^{+2.49}_{-0.03}$ &   $2.5^{+2.4}_{-1.3}$ &       $>60$ &      147.96 (156) \\
F2M1656+3821 & \ref{eqn:dpl}     & $1.83^{+0.22}_{-0.21}$ &  $0.9^{+1.6}_{-1.5}$ &      $1.1^{+1.2}_{-1.0}$ &      \ldots &      499.33 (495) \\
-            & \ref{eqn:mytorus} & $1.81^{+0.22}_{-0.20}$ &  $0.90^{+0.25}_{-0.01}$ &   $1.4^{+1.1}_{-0.8}$ &       $>60$ &      500.27 (494) \\
\enddata

\tablecomments{ The first row in each object's entry represents the phenomenological model fit parameters (Equation \ref{eqn:dpl}).  The second row represents the MYTorus model fit parameters (Equation \ref{eqn:mytorus}). }

\end{deluxetable}

Although we obtain good fits from an absorbed power law, the form of the model (Equation \ref{eqn:dpl}) is phenomenological and the different components may not necessarily be physically consistent with each other.  For example, an absorbed power law will result in reprocessed emission that is reliant upon the amount of absorption.  To better understand our spectra, a model is needed in which the intrinsic power law, absorption, Compton-scattering, and fluorescent line emission are treated self-consistently.  
Thus, we turn to MYTorus \citep{Murphy09}, which properly accounts for the processing of absorbed and Compton-scattered photons, as well as fluorescent line emission, to obtain a more physically-self-consistent model for these sources.

We follow the MYTorus model used by \citet{LaMassa16}, which includes an attenuated power law plus a Compton-scattered component and fluorescent line emission:

\begin{equation}
\begin{split}
{\rm model} = & ~C \times \exp[-N_{\rm H,Gal}\sigma(E)] \\
		& \times [A \times E_r^{-\Gamma} \times {\rm MYTorusZ}(N_{H,Z},\theta_{\rm obs},E_r) \\
		& + A_S \times {\rm MYTorusS}(A, \Gamma, N_{H,S},\theta_{\rm obs},E_r) \\
		& + A_L \times {\rm MYTorusL}(A, \Gamma, N_{H,S},\theta_{\rm obs},E_r) \\
		& + f_{\rm scatt} \times (A \times E_r^{-\Gamma})].\\
\label{eqn:mytorus}
\end{split}
\end{equation}

The files MYTorusZ, MYTorusS, and MYTorusL contain pre-calculated and tabulated parameters, that are used in the fitting.  They are derived via Monte Carlo calculations that take into account the reprocessing of the intrinsic AGN continuum in a toroidal structure for a range of column densities. MYTorusZ is the portion of intrinsic continuum that makes it through any absorbing or scattering medium along the line-of-sight (so-called `zeroth-order spectrum', {\em mytorus\_Ezero\_v00.fits}).  MYTorusS tabulates Compton-scattered emission that is added to the zeroth-order spectrum ({\em mytorus\_scatteredH500\_v00.fits}). MYTorusL provides tabulated fluorescent line emission that is also added to the zeroth-order spectrum ({\em mytl\_V000010nEp000H500\_v00.fits}), where {\em H500} refers to the termination energy of the model (500 keV).  
The chosen MYTorusZ and MYTorusL are the same as \citet{LaMassa16} used for the analysis of the two {\em NuSTAR}-observed red quasars.
MYTorusS differs slightly from the file used by \citet{LaMassa16}, whose termination energy was chosen to be 200 keV.  Since the high energy range of {\em XMM} is far below 200 keV (and, thus, 500 keV) the results are unaffected by this choice of MYTorusS. 
All three MYTorus components are needed in the modeling in order to preserve the self-consistency of the model.  
 
To preserve the self-consistency of the MYTorus model, $\Gamma$ and the power-law normalizations are tied among the MYTorus model components. Additionally, $N_{H,Z}$, $N_{H,s}$, the torus inclination angle, and the normalizations of the Compton-scattered and fluorescent-line-emission components ($A_S$ and $A_L$, respectively) are tied together, mimicking a torus with a homogenous distribution (i.e., the global column density is the same as the line-of-sight column density). 
We also add a separate power-law component, as we did in Equation \ref{eqn:dpl} to account for the soft excess seen below 1 keV.  This component, where the power law slope and normalization are linked to that of the intrinsic AGN continuum, is then treated as leaked emission that enters our line-of-sight directly. 

\begin{figure*}
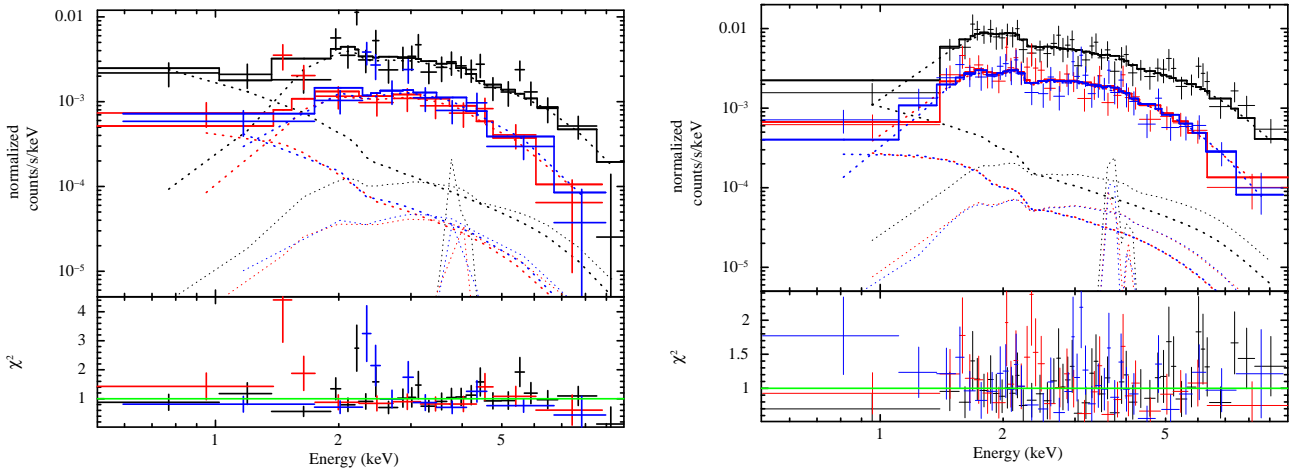

\includegraphics[angle=270,scale=0.35]{f4a.ps}
\includegraphics[angle=270,scale=0.35]{f4b.ps}
\caption{MYTorus plus scattered power law (Equation \ref{eqn:mytorus}) fit to the {\em XMM} quasar spectra in the energy range  $0.5 - 10$ keV. {\em Left --} F2M1113+1244. {\em Right --} F2M1656+3821.
The dotted lines show the model components and the solid line represents the combined model. Red and blue points and lines are for data from the MOS detectors. Black points and lines are for data from the PN detector.  Bottom panels show the residuals.
In both spectra the scattered or leaked emission dominates the fit below 1 keV while the zeroth-order spectrum (MYTorusZ) dominates above 1 keV, with the scattered (MYTorusS) and line emission (MYTorusL) contributing at a significantly lower level.}\label{fig:mytordpl}
\end{figure*}

For both sources, the MYTorus model parameters converge to nearly identical values as the phenomenological model in Equation \ref{eqn:dpl}.  The second row of each objects' entry in Table \ref{tab:param} lists in the best-fit parameters to the MYTorus model. The fitted spectra are shown in Figure \ref{fig:mytordpl}. Note that both the fluorescent-line-emitting component (thin dotted line that appears as a spike around 4 keV) and the Compton-scattered component (thin dotted line that spans 1-10 keV) are weak (contributing $\lesssim5\%$ of the zeroth-order emission beyond 1 keV) which provide strong constraints on the column density and inclination angle.  The normalization of this component, $f_{\rm scatt}$, provides the fraction of leaked radiation, which, at 2.5\% and 1.4\% for  F2M1113+1244 and F2M1656+3821, respectively, are consistent with the phenomenological model as well.

We present the soft ($0.5-2$ keV), hard ($2-10$ keV), and full ($0.5-10$ keV) X-ray fluxes of our quasars based on the MYTorus fits to our spectra in Table \ref{tab:fluxes}.  We also compute the intrinsic luminosities, corrected for absorption, scattering, and reflection, in Table \ref{tab:lumin}.  

\section{Discussion}




\begin{deluxetable}{cccc}



\tablewidth{0pt}

\tablecaption{Observed X-ray fluxes \label{tab:fluxes}}

\tablenum{4}

\tablehead{\colhead{Name} & \colhead{$F_{0.5-2 {\rm keV}}$} & \colhead{$F_{2-10 {\rm keV}}$} & \colhead{$F_{0.5-10 {\rm keV}}$} \\ 
\colhead{} & \colhead{($10^{-15}$ erg cm$^{-2}$ s$^{-1}$)} & \colhead{($10^{-13}$ erg cm$^{-2}$ s$^{-1}$)} & \colhead{($10^{-13}$ erg cm$^{-2}$ s$^{-1}$)} } 

\startdata
F2M1113+1244 &  $7.3^{+9.3}_{-4.0}$ &   $1.1^{+1.4}_{-0.6}$ & $1.2^{+1.5}_{-0.6}$ \\
F2M1656+3821 &  $13.3^{+7.6}_{-4.6}$ &  $2.0^{+1.2}_{-0.7}$ & $2.1^{+1.2}_{-0.8}$ \\
\enddata




\end{deluxetable}




\begin{deluxetable}{ccccc}



\tablewidth{0pt}
\tablecaption{Absorption corrected, Rest-frame X-ray and Infrared Luminosities \label{tab:lumin}}

\tablenum{5}

\tablehead{\colhead{Name} & \colhead{$\log L_{0.5-2 {\rm keV}}$} & \colhead{$\log L_{2-10 {\rm keV}}$} & \colhead{$\log L_{0.5-10 {\rm keV}}$} & \colhead{ $\log L_{6 \mu{\rm m}}$\tablenotemark{a} } \\ 
\colhead{} & \colhead{(erg s$^{-1}$)} & \colhead{(erg s$^{-1}$)} & \colhead{(erg s$^{-1}$)}  & \colhead{(erg s$^{-1}$)} } 
\startdata
F2M1113+1244 & $44.54^{+0.55}_{-0.24}$ & $44.56^{+0.55}_{-0.24}$ & $44.85^{+0.55}_{-0.24}$ & 46.6 \\
F2M1656+3821 & $44.54^{+0.25}_{-0.15}$ & $44.74^{+0.25}_{-0.15}$ & $44.95^{+0.25}_{-0.15}$ & 45.9 \\
\enddata
\tablenotetext{a}{The 6 $\mu$m luminosity was estimated by interpolating between the {\em WISE} photometric bands.}
\end{deluxetable}

\subsection{The $\Gamma - L/L_{\rm Edd}$ Relation}

\citet{Urrutia12} used SED modeling to determine the bolometric luminosities of the quasars in our sample. When combined with broad-line widths measured from our optical and near-infrared spectroscopy, their accretion rates are estimated and listed in Table \ref{tab:objects} as $\log(L/L_{\rm Edd})$. 

\begin{figure}
\plotone{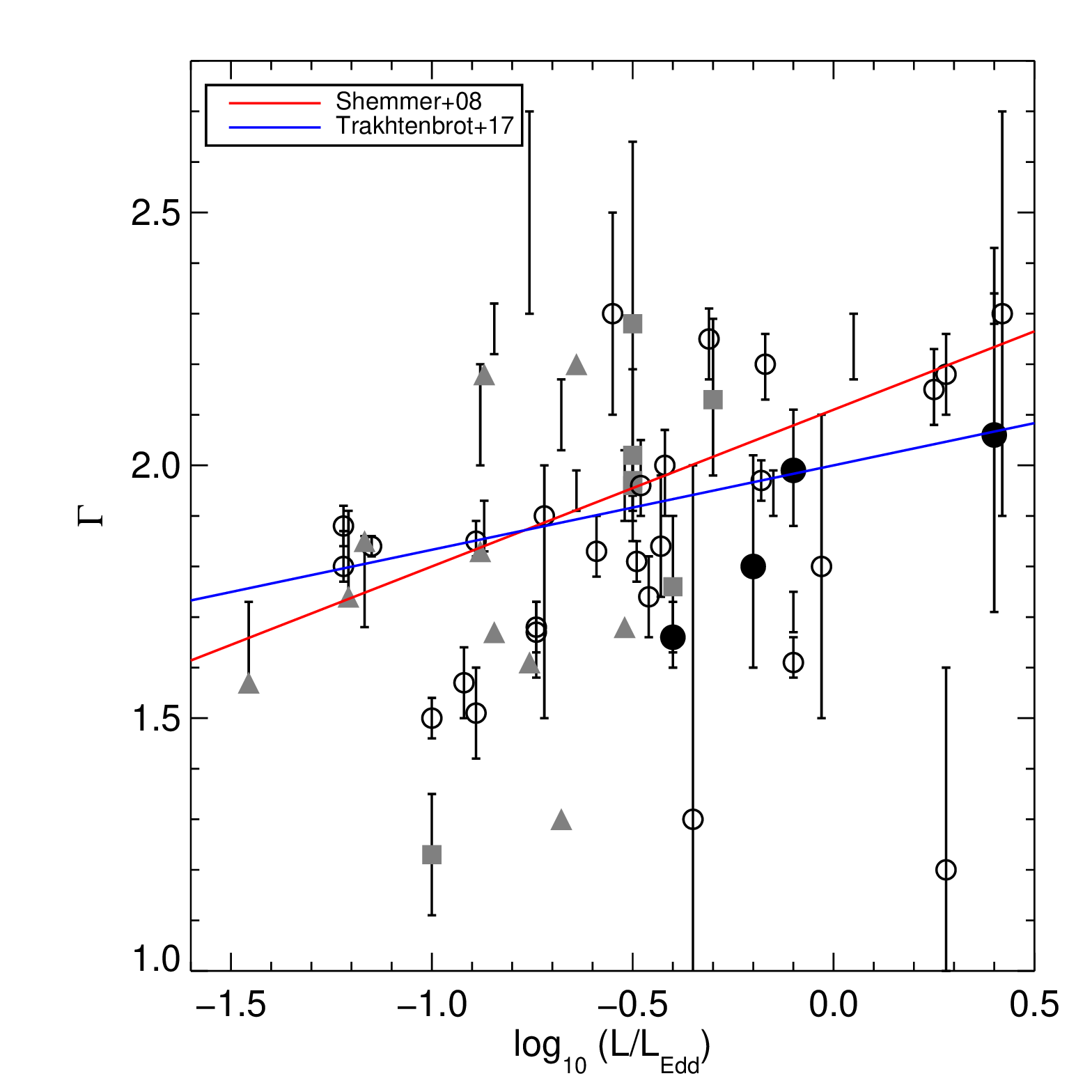}
\caption{X-ray photon index, $\Gamma$, versus Eddington ratio, $L/L_{\rm Edd}$ for luminous quasars.
Grey filled squares are luminous unobscured quasars from \citet{Shemmer06} and \citet{Shemmer08}.  Open circles are PG quasars with $\Gamma$ from \citet{Piconcelli05} and $L/L_{\rm Edd}$ from \citet{Inoue07}.  Grey filled triangles are local Compton thick AGN from \citet{Brightman16}.
Filled black circles are F2M quasars.  The red line is the $\Gamma - L/L_{\rm Edd}$ relation from \citet{Shemmer08}. The blue line is the \citet{Trakhtenbrot17} relation based on their BCES (Y|X) and FITEXY linear regression analysis methods. 
Red quasars appear to have flatter $\Gamma$'s relative to their optical/IR derived accretion rates, but obey the general trend of steeper $\Gamma$ with increasing accretion rate.}\label{fig:leddgamma}
\end{figure}

The X-ray photon index, $\Gamma$, has been shown to correlate with the BH accretion rate, $L/L_{\rm Edd}$, for high luminosity quasars ($44 \lesssim \log [\nu L_\nu(5100{\rm \AA})] \lesssim 48$; e.g., \citealt{Piconcelli05,Shemmer06,Shemmer08}).
In Figure \ref{fig:leddgamma}, we plot our derived $\Gamma$'s vs.~$L/L_{\rm Edd}$ along with luminous unabsorbed quasars from \citet{Piconcelli05}, \citet{Shemmer06}, and \citet{Shemmer08} as well as Compton-thick AGN from \citet{Brightman16}. The red quasars fall below the best-fit line from \citet{Shemmer08}, though they are largely consistent given the uncertainties on $\Gamma$.  It is unclear whether there is a systematic offset in the relation for red quasars given the small size of our sample. However, the trend of steepening photon index with increased accretion rate is apparent.  

We note that a recent study by \citet{Trakhtenbrot17}, using a large set of low-redshift AGN with high-quality optical and X-ray data spanning 0.3-195 keV \citep[{\em Swift}/BAT AGN;][]{Baumgartner13}, find weak to no correlation between the two quantities when $\Gamma$ is computed over the full available energy range. They attribute much of the original results to possibly oversimplified X-ray spectral modeling
and chosen method for determining $L_{\rm bol}$.  Thus, it may be that not much physical meaning can be extracted from the fitted photon index of red quasars' X-ray spectra in this parameter space. 

The blue line in Figure \ref{fig:leddgamma} shows the correlation derived by \citet{Trakhtenbrot17} using the full 0.3-195 keV energy range to determine $\Gamma$ and estimating $L_{\rm bol}$ from absorption-corrected X-ray luminosity.  The two red quasars that agree very well with this relation, F2M1113+1244 and F2M1227+3214, have lowest luminosities among the four F2M sources. The sample of {\em Swift}/BAT AGN used in the \citet{Trakhtenbrot17} study is dominated by objects with $L_{2-10 {\rm keV}} \sim 10^{42.5 - 44.5}$ and F2M1113+1244 and F2M1227+3214 fall within this range.  The other two red quasars, F2M1656+3821 and F2M0830+3506 are more luminous, yet have flatter photon indices, lower accretion rates, and disagrees with both relations.  

\subsection{X-ray Absorption vs. Dust Reddening}

\begin{figure}
\plotone{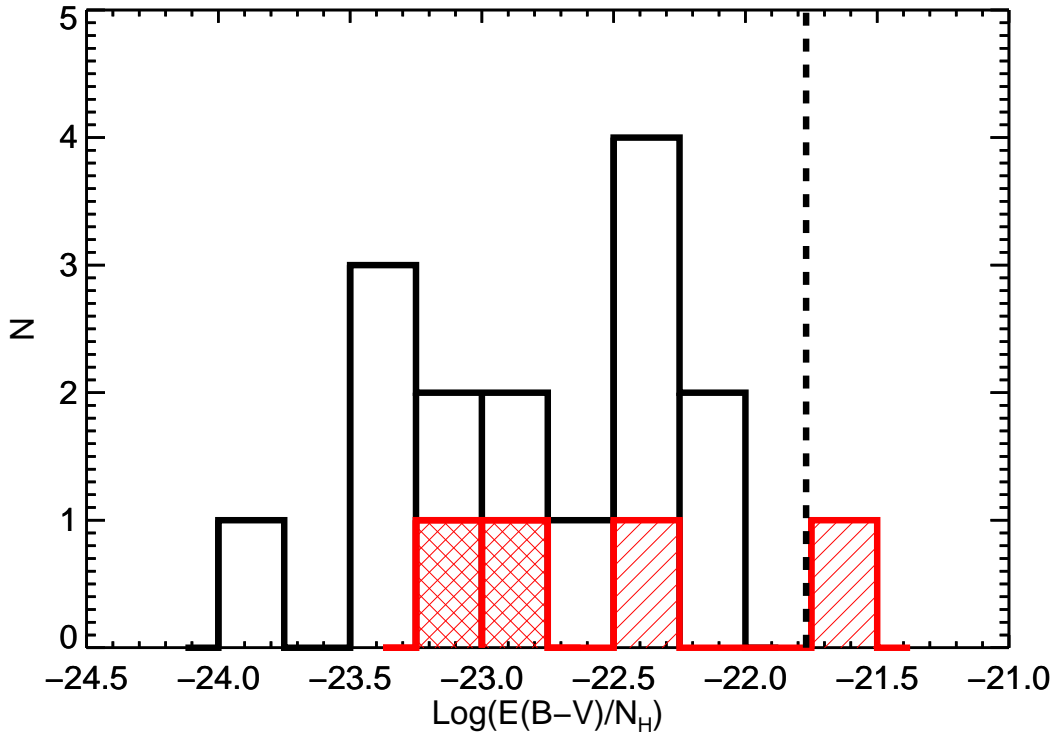}
\caption{Dust-to-gas ratios for the red quasars studied here (red cross hatched bins) and in \citet{LaMassa16} (red diagonal hatched bins). For comparison, we show the same for the sample studied by \citet{Maiolino01} of nearby AGN. The Milky Way dust-to-gas value is shown by the vertical dashed line.  The red quasars whose host galaxy morphologies show merger remnants all lie below the Galactic value, which can be explained by the absorbing gas and dust existing on different scales. }\label{fig:ebv_nh}
\end{figure}

We compute the dust-to-gas ratio for the two quasars so as to compare the optical reddening, $E(B-V)$, to their hydrogen column density, $N_H$. For F2M1113+1244 we find $\log(E(B-V)/N_H) = -22.95$ and for F2M1656+3821 $\log(E(B-V)/N_H) = -23.17$.  These values are very consistent with each other and are shown in Figure \ref{fig:ebv_nh} (red cross hatched bins) along with the two quasars, F2M0830+3506 and F2M1227+3214, studied by \citet{LaMassa16} (red diagonally filled bins). 

F2M1113+1244 and F2M1656+3821, studied here, as well as F2M0830+3506, have dust-to-gas ratios below the Milky Way value, and are distributed with the low redshift, moderate luminosity AGN studied by \citet{Maiolino01}.  These three sources all reside in merger remnants, according to their {\em HST} images.  The low dust-to-gas ratios could be explained by a model in which the gas that absorbs the X-rays and the dust that reddens the optical light are on different scales.  Since F2M quasars are all Type 1 objects, with broad emission lines atop a reddened continuum, our line of sight to the nucleus is out of the plane of accretion and likely does not intersect any dusty nuclear obscuring structure.  It is still possible that X-rays may be absorbed by atomic gas within the sublimation radius of the AGN, i.e., within the broad-line region, as depicted in the clumpy torus model of  \citet{Nenkova08}.  Outside the nuclear region, our line of sight may then intersect dust clouds in the host galaxy that moderately attenuates the optical spectrum as well as X-rays. 

F2M1227+3214 does not show any extended emission in its SDSS image, though its host galaxy may be outshone by the bright (while red) nuclear emission.  Thus the origin of this quasar's absorption is unclear.    

The dust-to-gas ratios for the two red quasars added in this study deviate significantly from the Galactic value, and from dust-to-gas ratios found in other reddened quasar samples.  The 2MASS-selected red AGN \citep{Cutri01} is another sample that has been studied in X-rays and across the electromagnetic spectrum.  While the AGN in this sample are generally at lower redshifts and luminosities as compared with the F2M sample, they can serve as instructive for comparison.  X-ray analyses of 2MASS red AGN find dust-to-gas ratios consistent with the Galactic value, albeit  typically lower by a factor of a few \citep{Wilkes05,Pounds07,Piconcelli10,Kuraszkiewicz09a}.  And \citet{Banerji14} find that the dust-to-gas ratios of a high redshift, heavily reddened quasar residing in a hyperluminous infrared galaxy (HyLIRG) is consistent with the Galactic ratio. \citet{Urrutia05} use hardness-ratios from short-exposure {\em Chandra} observations of twelve F2M quasars and estimate gas-to-dust ratios $\lesssim 20$ times higher than the Galactic value, implying {\em lower} dust-to-gas ratios. The two quasars in this paper, F2M1113+1244 and F2M1656+3821, have dust-to-gas ratios lower than the Galactic value by a factor of $\sim 15$ and $\sim 25$, respectively. 

\subsection{The $L_X - L_{IR}$ Relation}\label{sec:lxlir}

\begin{figure}
\plotone{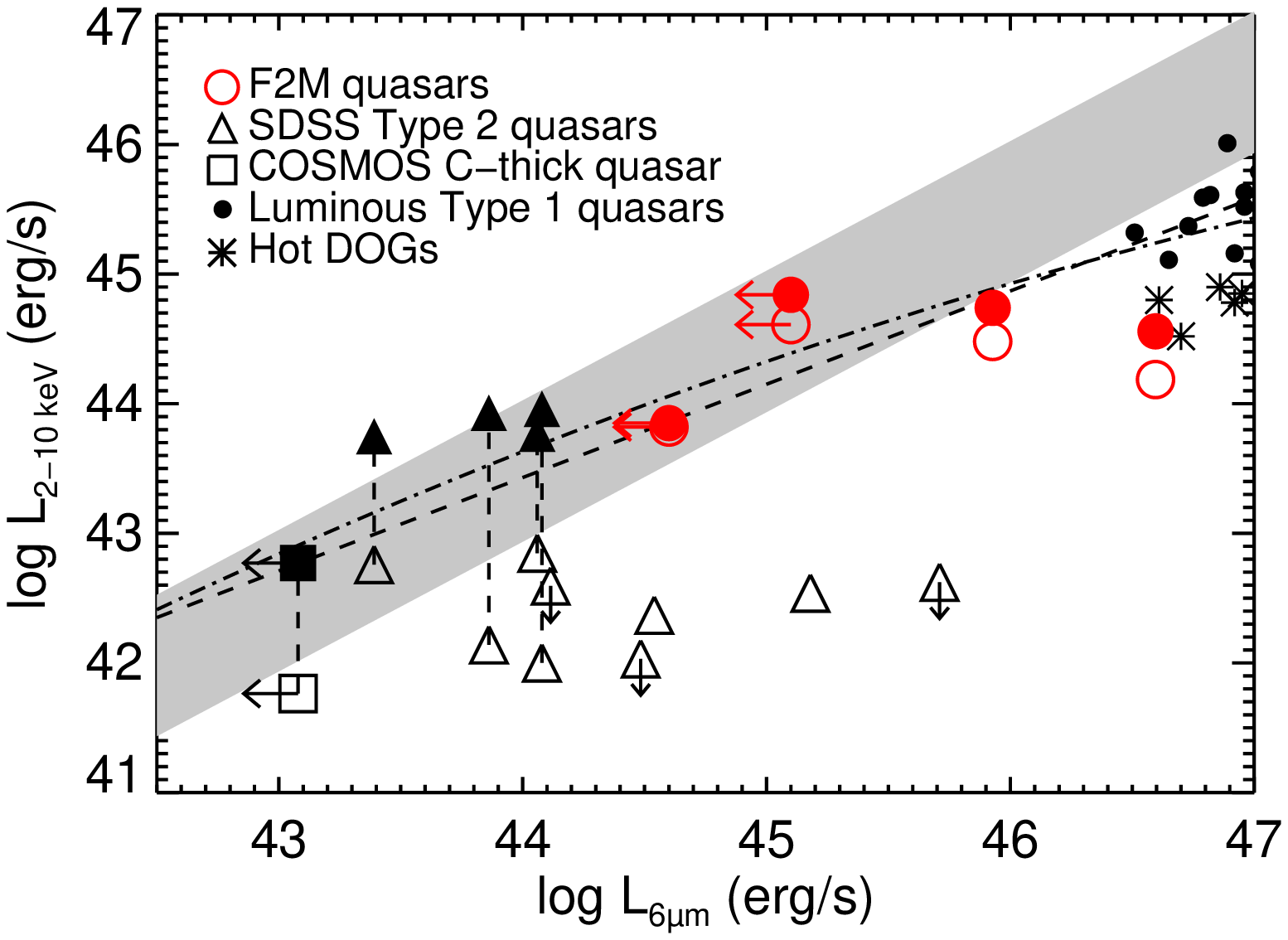}
\caption{Rest-frame 6 $\mu$m luminosity vs. rest-frame $2-10$ keV X-ray luminosity with no absorption correction (open symbols) and intrinsic, absorption-corrected X-ray luminosity (filled symbols).  The red circles are the F2M sources.  For comparison, we show SDSS Type 2 quasars (triangles) from \citet{Lansbury14,Lansbury15}, a Compton-thick {\em NuSTAR} quasar \citep[square;][]{Civano15}, and luminous, unobscured, Type 1 quasars (small black circles) from \citet{Stern15}.  Asterisks show Hot DOGs \citep{Ricci16}, which are infrared-hyperluminous, heavily obscured quasars.
 We also show X-ray to mid-infrared relations derived for various AGN samples in the literature.  The shaded region shows the \citet{Lutz04} relation derived from local Seyfert galaxies.  The dashed line represents the relation from \citet{Fiore09} derived from AGN in deep fields, while the dot-dashed line is the recently-derived relation from \citet{Stern15} that includes luminous quasars and that best-applies to the red quasars.   While the lower-luminosity red quasars are in general agreement with all three relations, the most IR-luminous source, F2M1113+1244, lies well below the relation.} \label{fig:lxlir}
\end{figure}

A relationship between X-ray and mid-infrared luminosity has been established for low-redshift and low-luminosity AGN \citep{Lutz04,Gandhi09}.
\citet{LaMassa16} showed that the 6 $\mu$m luminosities of F2M0830+3759 and F2M1227+3214 correlate exceptionally well with their X-ray derived values (within 0.1 dex of each other) in both their $2-10$ keV and $10-40$ keV luminosities, in agreement with the local relation.  Our {\em XMM-Newton} data allow us to determine whether F2M1113+1244 and F2M1656+3821 also obey these relations in the $2-10$ keV band.  Figure \ref{fig:lxlir} places these two additional red quasars on the relation \citep[shaded region;][]{Lutz04} shown in the left-hand panel of Figure 8 in \citet{LaMassa16} using the X-ray luminosities presented in Table \ref{tab:lumin}. The observed luminosities are shown with open symbols, while intrinsic, absorption-corrected luminosities are shown with filled symbols.  We determine the 6$\mu$m luminosity by interpolating between the infrared fluxes in the {\em WISE} bands.  

For comparison we plot previously-studied Type 2 quasars from SDSS \citep[triangles;][]{Lansbury14,Lansbury15}, a Compton-thick AGN identified in the COSMOS survey \citep[square;][]{Civano15}, and the set of luminous Type 1 quasars (small black circles) used by \citet{Stern15} to derive an X-ray to mid-IR relation that spans a luminosity range extending into the regime of highest luminosities.  This relation is plotted with a dot-dashed line.   Other relations, derived from samples at different luminosity and redshift regimes are also plotted.  The gray shaded area shows the relation from local Seyfert galaxies derived by \citet{Lutz04}. The dashed line is the relation from \citet{Fiore09} based on Type 1 AGNs in the COSMOS and CDF-S deep fields.  

Because the F2M red quasars are among the most intrinsically luminous quasars, the \citet{Stern15} relation is most applicable for comparison.  The three red quasars with the lowest infrared luminosities are in general agreement with the \citet{Stern15} relation.  
However, the red quasar with the highest infrared luminosity, F2M1113+1244, lies well below the relation and may be either underluminous in X-rays or, overluminous in the infrared.  The latter scenario may arise in a system with an abnormally large amount of hot circumnuclear dust.  The asterisks plotted in Figure \ref{fig:lxlir} show the location of hot dust-obscured galaxies (Hot DOGs) from \citet{Ricci16} that host some of the most luminous obscured quasars known, and are seen to lie below the \citet{Stern15} relation as well.  Hot DOGs are also considered transitional systems possibly representing a brief phase in some quasars' evolutions \citep{Piconcelli15,Fan16}.  
\citet{Ricci16} interpret the Hot DOGs as being underluminous in X-rays, rather than overluminous in the infrared.  This may be the case for F2M1113+1244 as well.  To see if red quasars and Hot DOGs form an independent X-ray to mid-infrared relationship specific to luminous, dusty, transitional systems, a larger sample of X-ray and mid infrared observations would be needed to span the luminosity range.

\section{Conclusions}

We presented {\em XMM-Newton} spectra for two luminous, dust-reddened F2M quasars at $z\sim 0.7$ that reside in host galaxies that are actively merging and whose multi-wavelength emission imply accretion rates near (or above) the Eddington limit.  Our exposure times resulted in well above 500 counts in both cases, which enabled robust spectral modeling. We find that their X-ray spectra are best fit by an absorbed power law with some leaked or scattered flux at the few-percent level.  The quasars are moderately absorbed with column densities of $N_H \sim 10^{23}$ cm$^{-2}$.  Their resultant dust-to-gas ratios are low compared with the Milky Way value, which, combined with the fact that their optical through near-infrared spectra show broad emission lines, imply a line of sight that does not intersect dust in an obscuring torus but rather dust in their host galaxies, while the X-ray absorption is by neutral gas inside the dust sublimation radius.  As galaxy mergers are associated with excess dust (e.g., LIRGs and ULIRGs) these findings are consistent with dusty mergers driving the black hole growth in these quasars.  

In addition, these red quasars are very luminous in the infrared (6 $\mu$m) and have X-ray luminosities that are low, compared with other luminous (unobscured) quasars at these wavelengths.  The two red quasars in this study are more akin to the class of extremely luminous, infrared-selected, heavily obscured quasars (so-called Hot DOGs), which also have a lower than expected X-ray luminosity given their even-higher luminosity in the infrared. 

In general we find that the X-ray properties of the red quasars studied in this paper, cannot yet be used as independent probes of underlying physical parameters, such as $L/L_{\rm Edd}$, $L_{\rm bol}$ (from the relation between X-ray and mid-infrared), or reddening, given their spread in dust-to-gas ratios.  Instead, we find that the X-ray properties of  red quasars do not obey normal AGN diagnostics and relations.  If independent relations exist for these transitional objects, a larger set of X-ray observations are needed to uncover them. 

\acknowledgments
We thank Daniel Stern and Benny Trakhtenbrot for helpful discussion.  
This work was supported by NASA under award No. NNX16AE17G.  EG acknowledges the generous support of the Cottrell College Award through the Research Corporation for Science Advancement. 

\facility{XMM (EPIC)}

\bibliographystyle{apj}
\bibliography{xmm_eg_astroph.bbl}

\end{document}